\documentclass[12pt,reqno]{amsart}
\usepackage{amsmath,amsfonts,amssymb,amscd,amsthm,amsbsy, color, marginnote}
\usepackage{graphicx}
\usepackage{subfig}
\newtheorem{theorem}{Theorem}
\newtheorem{meta-thm}[theorem]{Meta-Theorem}

\newtheorem{definition}[theorem]{Definition}

\newcommand\beq[1]{ \begin{equation}\label{#1} }
\newcommand{\eeq}{ \end{equation} }

\newcommand\beqa[1]{ \begin{eqnarray} \label{#1}}
\newcommand{\eeqa}{ \end{eqnarray} }
\newcommand{\beqano}{ \begin{eqnarray*} }
    \newcommand{\eeqano}{ \end{eqnarray*} }
\newcommand\equ[1]{{\rm (\ref{#1})}}

\def\integer{{\mathbb Z}}

\def\real{{\mathbb R}}
\def\torus{{\mathbb T}}

\title[{Nekhoroshev estimates for the orbital stability of Earth's satellites}]
{Nekhoroshev estimates\\ for the orbital stability of Earth's satellites}

\author{Alessandra Celletti, Irene De Blasi, Christos Efthymiopoulos}
\date{}

\begin{document}
\maketitle

\baselineskip=18pt

\begin{abstract}
We provide stability estimates, obtained by implementing the
Nekhoroshev theorem, in reference to the orbital motion of a small
body (satellite or space debris) around the Earth. We consider a
Hamiltonian model, averaged over fast angles, including the $J_2$
geopotential term as well as third-body perturbations due to Sun
and Moon. We discuss how to bring the Hamiltonian into a form
suitable for the implementation of the Nekhoroshev theorem in the
version given by \cite{Poschel} for the `non-resonant' regime. The
manipulation of the Hamiltonian includes i) averaging over fast
angles, ii) a suitable expansion around reference values for the
orbit's eccentricity and inclination, and iii) a preliminary
normalization allowing to eliminate particular terms whose
existence is due to the non-zero inclination of the invariant
plane of secular motions known as the `Laplace plane'. After
bringing the Hamiltonian to a suitable form, we examine the domain
of applicability of the theorem in the action space, translating
the result in the space of physical elements. We find that the
necessary conditions for the theorem to hold are fulfilled in some
non-zero measure domains in the eccentricity and inclination plane
$(e,i)$ for a body's orbital altitude (semi-major axis) up to
about 20\,000 km. For altitudes around 11\,000 km we obtain
stability times of the order of several thousands of years in
domains covering nearly all eccentricities and inclinations of
interest in applications of the satellite problem, except for
narrow zones around some so-called `inclination-dependent'
resonances. On the other hand, the domains of Nekhoroshev
stability recovered by the present method shrink in size as the
semi-major axis $a$ increases (and the corresponding Nekhoroshev
times reduce to hundreds of years), while the stability domains
practically all vanish for $a>20 000$ km.
\end{abstract}

\keywords{KEYWORDS. Stability, Nekhoroshev theorem, Resonance, Normal form,
Satellite, Space debris}


\section{Introduction}
\label{sec intro}
The study of the stability of the motion of celestial bodies is
relevant from both the theoretical and practical points of view;
such investigation can be approached using numerical or analytical
tools (see \cite{celletti2010stability}) for a review). In this
work, we consider the problem of the long-term (over $10^3-10^4$
years) stability of a small body (satellite or space debris) in
orbit around the Earth and subject to third-body perturbations due
to the Moon and the Sun. By stability we mean that the body undergoes no
large variations of its orbital elements that could produce a
drastic change (e.g., escape) in the orbit.

In the orbital study of satellite motions, it is convenient to split the space environment
around the Earth into three distinct regions according to the distance from the Earth's
surface, where different elements can affect the dynamics:
\begin{itemize}
\item
[$(i)$] LEO: Low-Earth-Orbit (from $90$ to $2\,000$ km of altitude), where the Earth's
atmosphere generates dissipative effects;
\item
[$(ii)$-$(iii)$] MEO: Medium-Earth-Orbit (between $2\,000$ and
$30\,000$ km of altitude) and GEO: Geostationary-Earth-Orbit
(altitudes around the geosynchronous orbit at about $35\,786$ km),
where the dissipative effect of the atmosphere is negligible and
the dynamical system associated with the equations of motion is
conservative.  In these regimes, the most important contributions
are due to the geopotential and to the lunar and solar third-body
perturbations.
\end{itemize}

We will hereafter consider a Hamiltonian model for the motion of
small bodies at MEO (see, instead,
\cite{celletti2018dynamics,Lho2016} for the inclusion of
dissipative effects). The study of dynamics at MEO in the
conservative regime has been subject of many works, including the
development of analytical models (e.g.,
\cite{CGmajor,celletti2017dynamical,giacaglia,Kaula1962,Lane1989}),
study of resonances (e.g.,
\cite{sB01,breiter2001coupling,CGL2020,CGPSIADS,CGPR2017,chaogick,
cook1962luni,EH,Hughes,LDV}), as well as the dynamical
chartography (stability maps, onset of chaos) of the MEO region
(e.g.,
\cite{ADRRVDQM,CPL2015,daquin2016dynamical,gkolias2016order,Rosengren2013,aR15,
rosengren2016galileo,rossi2008resonant,skoulidou2019medium,VL}).

The aim of this work is to study the stability of a model for
objects in MEO from an analytical point of view,
providing exponential stability estimates using the celebrated
\textit{Nekhoroshev theorem} (\cite{nekhoroshev1977exponential}).
We stress that, while the Nekhoroshev theorem is particularly
relevant for systems with three or more degrees of freedom, which
can be affected by the phenomenon known as \textit{Arnold diffusion} (\cite{Arnold64}), the applicability of the theorem in
securing the long-term stability in \textit{open domains} in the
action space holds for systems of any number of degrees of freedom
larger than or equal to two. Furthermore, the Nekhoroshev theorem
was originally developed under a suitable non-degeneracy
condition, called \textit{steepness}, while later approaches (e.g.
\cite{benettin1986stability, Poschel}) focus on the important
subcase of convex and quasi-convex Hamiltonians (see
\cite{Poschel} for definitions). As regards the applications, the
theorem was proved useful in obtaining realistic estimates of the
domains or times of practical stability of the orbits in a number
of interesting problems in Celestial Mechanics. Among others, we
mention the three-body problem (\cite{CF}) as well as the problem
of the Trojan asteroids (\cite{CG}, \cite{G2}).

In this work, we apply the Nekhoroshev theorem to a model approximating the (averaged
over short period terms) dynamics of a small body around the Earth. As discussed below,
this allows to obtain long-time stability estimates for realistic sets of parameters,
at least for altitudes (values of the semi major axis) below 20\,000 km.

We consider a model ruled by a Hamiltonian function obtained as
the sum of different contributions, namely, the geopotential
$J_2$ term as well as the third-body perturbations on the small
body by the Sun and Moon. Several studies (see
\cite{daquin2016dynamical, gkolias2016order,Aristoff2021,Nie2021} and references
therein) have demonstrated the relevance of this model in
capturing all important effects for the long-term dynamics at MEO.

An important aspect of our present work concerns a number of
preliminary operations performed on the initial Hamiltonian, which
turn to be crucial to the purpose of bringing the Hamiltonian in a
form allowing to explicitly demonstrate the fulfilment of the
conditions for the holding of the Nekhoroshev theorem in the form
given in \cite{Poschel}. These preliminary steps are explained in
detail in Section 2 below, and can be summarized as follows:

(i) \textit{Average over fast angles.} We start by averaging the
Hamiltonian over the problem's fast angles, i.e., the mean
anomalies of the small body's, Moon's and Sun's orbits. After this
operation, the semi-major axis $a$ of any orbit becomes a constant
which can be used to label the altitude of each orbit.

(ii) \textit{Expansion around reference values in the eccentricity and inclination}.
The remaining elements (eccentricity $e$ and inclination $i$),
which can be mapped into the action variables of the problem, undergo `secular' (slow)
evolution under the averaged Hamiltonian. Our purpose is to characterize the stability
of the orbits in the space $(e,i)$ of the orbital elements. To this end, fixing a grid
of reference values $(e_*,i_*)$ in the plane $(e,i)$ for each (constant) semi-major
axis $a=a_*$, we perform a Taylor expansion of the averaged Hamiltonian around the points
in action space associated with the reference point $(e_*,i_*)$. This step is important,
since the Taylor-expanded Hamiltonian can be easily manipulated in terms of normalizing
canonical transformations necessary to perform with the aid of a computer-algebraic
program (see below).

(iii) \textit{Preliminary normalization}. We perform a preliminary
normalization of the averaged and Taylor-expanded Hamiltonian,
aiming to eliminate some terms which, albeit reflecting a trivial
dynamics (see Section 3), may artificially affect the estimates
found by implementing P\"{o}schel's version of the Nekhoroshev
theorem.  We argue below that this step is a consequence of the
non-zero value of the inclination of the Laplace plane with
respect to the Earth's equatorial plane. The inclination can be
expressed as 
\begin{equation}\label{ip}
i^{(p)}\simeq-\frac{A}{2B}\frac{1}{(\mu_Ea)^{1/4}}\ ,
\end{equation}
where
\begin{equation}\label{A1,B1}
\begin{split}
      &A=-\frac{3R_E^2a^{7/4}\sin{(2i_0)}}{8(\mu_E)^{1/4}}
      \left(\frac{\mu_M}{a_M^3}+\frac{\mu_{\odot}}{a_{\odot}^3}\right),\\
      &B=\frac{3}{4}\frac{\sqrt{\mu_E}R_E^2J_2}{a^{7/2}}
      +\frac{3\mu_M(2-3\sin^2{i_0})}{16\sqrt{\frac{\mu_E}{a^3}}a_M^3}
      +\frac{3\mu_{\odot}(2-3\sin^2{i_0})}{16\sqrt{\frac{\mu_E}{a^3}}a_S^3}
\end{split}
\end{equation}
with $R_E, \mu_E$ being the Earth's mean radius and mass parameter, $a_M, \mu_M$
are the Moon's semimajor axis and mass parameter, $a_\odot, \mu_\odot$ are the Sun's
semimajor axis and mass parameter, and finally $i_0$ is the inclination of the
ecliptic plane. A key result in the present paper is the use of normal form techniques
to reduce the size, in the Hamiltonian, of all terms related to the Laplace plane
(see Section~\ref{sec prel}); whenever convergent, this procedure is crucial to
put the initial Hamiltonian in a form for which Nekhoroshev's nonresonant stability
estimates can be produced, since it allows us to control the norm of the perturbing
function under a suitable choice of the domain in the actions.

\begin{table}
\begin{center}
\begin{tabular}{|ccc|ccc|ccc|ccc|}
\hline
 $\alpha$  &$\beta$  &$i(deg)$  &$\alpha$  &$\beta$  &$i(deg)$
&$\alpha$  &$\beta$  &$i(deg)$  &$\alpha$  &$\beta$  &$i(deg)$ \\
\hline
 1         &0        &63.43     &0         &1        &90
&1         &1        &46.37     &1         &-1       &73.1     \\
 2         &1        &56        &1         &2        &0
&2         &-1       &69        &-1        &12       &78       \\
 3         &1        &58.75     &3         &-1       &67.33
&1         &-3       &81.47     &          &         &         \\
\hline
\end{tabular}
\caption{Inclination-dependent resonances of order $\leq4$ in the lunisolar model.
The coefficients are such that $\alpha\dot{\omega}+\beta\dot{\Omega}=0$, where $\omega$
is the argument of the perigee and $\Omega$ is the longitude of the ascending node
(\cite{Hughes}).}
\label{introtab}
\end{center}
\end{table}

Now, following steps (i) to (iii) above, the procedure leads to a
normalized 2-dimensional Hamiltonian expressed in suitably defined
action-angle variables $(\mathbf{I},
\mathbf{u})\in\real^2\times\torus^2$, of the form:
\begin{equation}\label{splittedH}
\mathcal{H}(\mathbf{I},
\mathbf{u})=h_{0}(\mathbf{I})+h_1(\mathbf{I},\mathbf{u})\ .
\end{equation}
Using the Hamiltonian \equ{splittedH}, we can derive stability
results on the eccentricity and the inclination by implementing
the estimates provided in Proposition 1 of \cite{Poschel}. This
proposition refers to the so-called \textit{non-resonant} regime,
i.e., when the fundamental frequencies deduced by the integrable
part of the Hamiltonian, $h_0$, are subject to no resonance
conditions.  Under particular assumptions on the non-resonance
condition for $h_0$, as well as on the smallness of the norm of
$h_1$ in a suitable functional space and domain in the action
variables (see Section \ref{ssec posch}), one can prove that the
actions remain in a small neighborhood of their initial values for
a period of time which is exponentially long with respect to the
norm of $h_1$. We remark that Proposition 1 in \cite{Poschel} does
not require any convexity assumption on the Hamiltonian. This
assumption is relevant when analyzing resonant regimes (for a
thorough analysis of different non-degeneracy conditions such as
convexity, quasi-convexity, 3-jet, etc., see \cite{DCE}). However,
we also stress that, despite our use of P\"{o}schel's proposition
in the non-resonant regime, the presence of resonances at MEO
plays an important role also in our results, as becomes evident in
the discussion of our results in Section 4. In fact, we find that
our obtained stability domains typically exclude some zones around
the so-called \textit{inclination-dependent resonances}
(\cite{Hughes}), i.e., resonances appearing for particular values
of the inclination of the orbit, independently of the value of the
semimajor axis or the eccentricity. This is because the series
constructed in our preliminary normalization of the Hamiltonian
are affected by small divisors related to the most-important of
these resonances, given in Table~\ref{introtab}. Also, the
frequencies associated with these divisors influence the
determination of the so-called `Fourier cut-off' (Section 3) which
appears in the implementation of the Proposition 1 of
\cite{Poschel}.

As described in Section \ref{ssec numres}, the stability estimates
obtained in this work strongly depend on the distance of the small
body from the Earth's center: our results show that the domain of
Nekhoroshev stability in the plane $(e,i)$ has a large volume
(limited only by narrow strips around resonances) at the distance
of 10\,000 km, while it shrinks to a near-zero volume beyond the
distance of 20\,000 km.  We should stress that this result is
partly due to the dynamics itself (the $J_2$ dynamics alone is
integrable, but the third body perturbations increase in relative
size as the distance from the Earth increases), but also probably
due, in part, to our particular technique used to apply the
Nekhoroshev theorem, i.e. including the processing of the
Hamiltonian as described in steps (i)-(iii) above. We thus leave
open the possibility that this latter constraint be relaxed with
the use of a better technique. Also, our present treatment is
simplified in that we ignore the periodic oscillation of the Moon's line
of nodes (by an amplitude of $~11.5^\circ$ over a period of $18.6$
yr) and inclination (by $\pm 5^\circ$) around the ecliptic of the
Moon's orbit with respect to the Earth's equatorial plane. This
oscillation introduces one more secular frequency to the problem;
however, it substantially affects the orbits only for semi-major
axes $a>20\,000$ km, which is, anyway, beyond the domain of
stability presently found even while ignoring this effect.

As a final remark, in \cite{DCE} we studied the satellite's stability in
two different models: the $J_2$ approximation of the geopotential and a model
that includes $J_2$,  Sun and Moon. We computed suitable normal forms and obtained stability results
by estimating the size of the remainder function after the normalizing transformation.
In \cite{DCE} we only considered values of the eccentricity and inclination
close to zero and to the inclination of the Laplace plane, respectively; besides, our stability estimates were not obtained separately for eccentricity and inclination, but only as regards their combination $\sqrt{1-e^2}\cos{i}$ through the Lidov-Kozai integral. This is a main difference with respect to the present paper, in which
we consider generic values for the eccentricity and inclination.

This article is organized as follows: Section \ref{sec ham} provides the construction
of the secular Hamiltonian function which describes the model, as well as its
normalization; in Section \ref{sec prel} the theoretical framework is described,
along with the algorithm used to produce the normalized Hamiltonian and the stability
estimates; Section \ref{ssec numres} describes the results.

\section{Hamiltonian preparation}
\label{sec ham}

In this Section, we provide details on the model (Section~\ref{sec:model}),
on the corresponding secular Hamiltonian function averaged over fast angles
(Section~\ref{ssec hamave}), the expansion around some reference values
for the eccentricity and inclination (Section~\ref{ssec hamexpei}), and the preliminary normalization to remove specific terms (Section~\ref{ssec norm}).

\subsection{Model}\label{sec:model}
We consider a small body (satellite or debris) $S$ of
infinitesimal mass, under the action of the Earth's gravitational
field and the third-body perturbations due to the Moon and Sun.
Geocentric inertial Cartesian coordinates are denoted by
$(x,y,z)$, where the $xy$-plane coincides with the Earth's
equatorial plane, $z$ points to the north pole, and $x$ points to
a fixed direction (ascending node of the Sun's geocentric orbit).
We denote by $\mathbf{r}(t)=(x(t),y(t),z(t))$ the time-evolving
radius vector of the body $S$, and by $(a,e,i,M,\omega,\Omega)$
the osculating orbital elements of $S$, where $a$ is the semimajor
axis, $e$ the eccentricity, $i$ the inclination with respect to
the Earth's equatorial plane, $M$ is the mean anomaly, $\omega$
the argument of the perigee and $\Omega$ the longitude of the
ascending node.

In the sequel, we consider the following approximation to the body's equations of motion:
\begin{equation}\label{eqmor}
\ddot{\mathbf{r}} = -\nabla V_E(\mathbf{r})
-\mu_{\odot}
\bigg(\frac{\mathbf{r}-\mathbf{r}_{\odot}}{|\mathbf{r}-\mathbf{r}_{\odot}|^3}
+\frac{\mathbf{r}_\odot}{\vert\mathbf{r}_\odot\vert^3}\bigg)-\mu_{M}
\bigg(\frac{\mathbf{r}-\mathbf{r}_{M}}{|\mathbf{r}-\mathbf{r}_{M}|^3}
+\frac{\mathbf{r}_M}{\vert\mathbf{r}_M\vert^3}\bigg)\ ,
\end{equation}
where $V_E(\mathbf{r})$ approximates the geopotential via the relation
\begin{equation}\label{potearth}
V_E(\mathbf{r}) = V_{kep}(|\mathbf{r}|)+V_{J_2}(\mathbf{r})\ ,
\end{equation}
where $V_{kep}(r)=-{\mu_E\over r}$ and $V_{J_2}$ in spherical
co-ordinates $(r,\varphi,\phi)$ is given by\
\begin{equation}\label{pottrunc}
V_{J_2}(r,\varphi,\phi)=
\frac{\mu_EJ_2}{r}\left\{\left(\frac{R_E}{r}\right)^2
\left(\frac{3}{2}\sin^2{\phi}-\frac{1}{2}\right)\right\}.
\end{equation}
In the above formulas:
\begin{itemize}
\item
[-] $\mathcal{G}$ is the gravitational constant,
$\mu_E=\mathcal{G}m_E$, $\mu_M=\mathcal{G}m_M$,
$\mu_{\odot}=\mathcal{G}m_{\odot}$ with $m_E$, $m_M$, $m_\odot$
the masses of the Earth, Moon and Sun respectively.
\item
[-] We adopt the value $J_2=1.082\times 10^{-3}$ for the $J_2$ coefficient, and
$R_E=6\,400$~km for the Earth's equatorial radius.
\item
[-] $\mathbf{r}$, $\mathbf{r}_{\odot}$ and $\mathbf{r}_M$ are, respectively, the
geocentric position vectors of $S$, Sun and Moon.
\end{itemize}
The expressions of $\mathbf{r}_{\odot}$ and $\mathbf{r}_M$ depend
on the assumptions on the orbits of Sun and Moon. In this work,
the geocentric orbit of the Sun is taken as a fixed ellipse with 
$a_{\odot}=1.496\times 10^8$ km, $e_{\odot}=0.0167$ and $i_{\odot}=23.44^{\circ}$,
while the geocentric orbit of the Moon is taken as a fixed ellipse
with orbital parameters $a_M=384\,748$ km, $e_M=0.0554$ and
$i_M=i_{\odot}$. The last assumption has an important effect on the
dynamics: it implies that the only lunisolar resonances which
affect the dynamics of the body are those whose location in the
element space $(a,e,i)$ depends only on the inclination (see
\cite{Hughes}). More resonances, instead, appear when the effect
of nodal precession (by a period of 18.6 years) of the Moon's
orbit is taken into account. However, these resonances affect the
dynamics only at altitudes exceeding the ones where we presently
establish Nekhoroshev stability (see \cite{gkolias2016order} and
Section 4 below), thus they can be ignored in the framework of our
present study.

The Hamiltonian function which describes the motion of $S$ can be expressed as the sum
of three contributions:
\begin{equation}\label{haminiz}
\mathcal{H}=\mathcal{H}_{E}+\mathcal{H}_{\odot}+\mathcal{H}_M,
\end{equation}
where $\mathcal{H}_E = \mathbf{p}^2/2 +V_E(\mathbf{r})$ with
$\mathbf{p}= \mathbf{\dot r}$, and $\mathcal{H}_{\odot}$ and
$\mathcal{H}_M$ are the solar and lunar third-body perturbation
terms. Considering the quadrupolar expansion of the third-body
perturbation terms in the equations of motion (\ref{eqmor}), we
find
\begin{equation}\label{bspot}
\begin{split}
  \mathcal{H}_\odot&=V_\odot(\mathbf{r})=
  -\frac{\mu_\odot}{|\mathbf{r}-\mathbf{r_\odot}|}
  +\frac{\mu_\odot}{r_\odot^3}\mathbf{r}\cdot\mathbf{r_\odot}\\
&=-\frac{\mu_\odot}{r_\odot}-\frac{\mu_\odot}{2r_\odot^3}r^2+\frac{3}{2}
\frac{\mu_\odot(\mathbf{r}\cdot\mathbf{r}_\odot)^2}{r_\odot^5}
+O\left(\left(\frac{r}{r_\odot}\right)^3\right)\ ,
\end{split}
\end{equation}
\begin{equation}\label{bmpot}
\begin{split}
  \mathcal{H}_M&=V_M(\mathbf{r})=
  -\frac{\mu_M}{|\mathbf{r}-\mathbf{r_M}|}
  +\frac{\mu_M}{r_M^3}\mathbf{r}\cdot\mathbf{r_M}\\
&=-\frac{\mu_M}{r_M}-\frac{\mu_M}{2r_M^3}r^2+\frac{3}{2}
\frac{\mu_M(\mathbf{r}\cdot\mathbf{r}_M)^2}{r_M^5}
+O\left(\left(\frac{r}{r_M}\right)^3\right)\ .
\end{split}
\end{equation}

\subsection{Average over fast angles - Secular Hamiltonian}
\label{ssec hamave}
The secular motion of the body $S$ can be modeled by computing the average
of (\ref{haminiz}) over all canonical angles associated to the fast
motions of $S$, the Sun and the Moon. Note that the period
of the Sun is only `semi-fast' (one year, compared to secular
periods of $\sim 10$ yrs for the small body), and more detailed
models can consider also the case of `semi-secular' resonances,
i.e., resonances in the case in which the equations of motion (and
Hamiltonian) are not averaged with respect to the Sun's mean
anomaly (see, for example, \cite{celletti2017dynamical}).

Averaging with respect to all fast angles leads to the following, called hereafter,
\textit{secular Hamiltonian}, given by the sum of the averaged contributions of the
Earth, Sun and Moon:
\begin{equation}\label{Hsec}
  \mathcal{H}^{(sec)}=
  \mathcal{H}_{E}^{(av)}+\mathcal{H}_{\odot}^{(av)}+\mathcal{H}_{M}^{(av)}\
.\end{equation}
The function $\mathcal{H}^{(sec)}=\mathcal{H}^{(sec)}(G,\Theta,\omega,\Omega)$ is a
two degrees of freedom Hamiltonian, which can be explicitly computed in terms of
Delaunay canonical action-angle variables $G$, $\Theta$ (with conjugated angles
$\omega$, $\Omega$), related to the orbital elements by the expressions (see, e.g.,
\cite{celletti2010stability}):
\begin{equation}
G=\sqrt{\mu_E a(1-e^2)}, ~~~\Theta=\sqrt{\mu_E a(1-e^2)}\cos i~~.
\end{equation}
Since the averaged Hamiltonian does not depend on the mean anomaly
$M$ of $S$, the conjugated Delaunay action $L=\sqrt{\mu_E a}$, and
hence the semi-major axis $a$, is a constant of motion of the
Hamiltonian $\mathcal{H}^{(sec)}$. We set $L=L_*$, or,
equivalently, $a=a_*$ when referring to trajectories whose
semi-major axis has the reference value $a_*$.

Following a well-known procedure (e.g., \cite{Kaula1962}), the various terms in the
secular Hamiltonian can be computed as follows: for the geopotential term we have
$$
\mathcal{H}_E^{(av)}=
\frac{1}{2\pi}\int_0^{2\pi}(\mathcal{H}_{kep}+\mathcal{H}_{J_2})dM
=\mathcal{H}_{kep}^{(av)}+\mathcal{H}_{J_2}^{(av)}\ ,
$$
where $\mathcal{H}_{kep}=\mathbf{p}^2/2 +V_{kep}$,
$\mathcal{H}_{J_2}=V_{J_2}$, which leads to 
\begin{equation}\label{HJ2avfin}
  \mathcal{H}_E^{(av)}
  =
  -\frac{\mu_E^2}{2L^2}
  -J_2\frac{\mu_E R_E^{2}}{a_*^3(1-e^2)^{3/2}}
  \left(\frac{1}{2}-\frac{3}{4}\sin^2{i}\right)\ .
\end{equation}
For the terms $\mathcal{H}_{\odot}^{(av)}$, $\mathcal{H}_{M}^{(av)}$ we compute
the integral
$$
\mathcal{H}_{\odot}^{(av)} = {1\over 4\pi^2}
\int_0^{2\pi}\int_0^{2\pi}
\left(
-\frac{\mu_\odot}{r_\odot}-\frac{\mu_\odot}{2r_\odot^3}r^2+\frac{3}{2}
\frac{\mu_\odot(\mathbf{r}\cdot\mathbf{r}_\odot)^2}{r_\odot^5}
\right) dM dM_\odot
$$
(and analogously for $\mathcal{H}_{M}^{(av)}$); it is convenient
to change the integration variables from $M$ to $u$ (eccentric
anomaly of $S$) and from $M_\odot$ to $f_\odot$ (true anomaly of
the Sun). We note that, up to quadrupolar terms, this yields the
same result as considering the Moon and Sun in circular, instead
of elliptic, orbits (in which case $M_\odot$, $M_M$ would be equal
to $f_\odot, f_M$), but replacing each third-body's semi-major
axis $a_b$ with the expression $a_b\rightarrow a_b(1-e_b^2)^{1/2}$
(index $b$ standing for Sun or Moon). This replacement
accomplishes the first step in the Hamiltonian preparation.

\subsection{Expansion around reference values $(e_*,i_*)$}
\label{ssec hamexpei}
After performing the above operations, the Hamiltonian
$\mathcal{H}^{(sec)}$ becomes a function of the body's
action-angle variables $(G,\omega)$, $(\Theta,\Omega)$, while it
depends also on the Delaunay action $L$, which however, does not
affect the secular dynamics and can be carried on all subsequent
expressions as a parameter (equal to $L_*$). We use,
alternatively, $a_*$ as the parameter appearing in the
coefficients of all trigonometric terms in $\mathcal{H}^{(sec)}$.
Furthermore, it turns convenient to express $\mathcal{H}^{(sec)}$
in terms of \textit{modified Delaunay variables} instead of the
original Delaunay variables. Let $\delta L=L-L_*$ with
$L_*=\sqrt{\mu_E a_*}$. We employ the modified Delaunay variables
$(\delta L,\Gamma,\tilde{\Theta}, \lambda, p, q)$, related to the
original Delaunay variables $(L, G, \Theta, M, \omega, \Omega)$
via the latters' dependence on the Keplerian
elements $(a,e,i,M,\omega,\Omega)$. We have 
\begin{equation} \label{moddel}
\begin{cases}\delta L=L-L_*
=\sqrt{\mu_E a}-\sqrt{\mu_E a_*}\\ \Gamma=L-G
=\sqrt{\mu_E a}(1-\sqrt{1-e^2})\\
\tilde{\Theta}=G-\Theta
=\sqrt{\mu_E a}\sqrt{1-e^2}(1-\cos{i})\\
\end{cases}
\begin{cases}
\lambda=M+\omega+\Omega\\p=-\omega-\Omega\\
q=-\Omega\ . \\
\end{cases}
\end{equation}

Starting now from the Hamiltonian
$\mathcal{H}^{(sec)}(\Gamma,\tilde{\Theta},p,q)$, our goal will be
to examine Nekhoroshev stability in a covering of the action space
in terms of local neighborhoods around a grid of reference values
corresponding to a grid of element values  $(a_*,e_*,i_*)$ (see
Section~\ref{ssec alg}). This motivates to introduce the variables
$P'$ and $Q'$ defined by
\begin{equation}
\begin{cases}
P=\Gamma_*-\Gamma, \\
Q=\tilde{\Theta}_*-\tilde{\Theta}\ ,
\end{cases}
\end{equation}
where $\Gamma_*$ and $\tilde\Theta_*$ are the values corresponding
to the orbital elements $(e_*,i_*)$, and compute the Taylor
expansion of $\mathcal{H}^{(sec)}$ in powers of the small
quantities $(Q,P)$, truncated at a maximum order $N$ (we set
$N=12$). We then arrive at the following truncated secular
Hamiltonian model
\begin{equation}\label{hamsum}
\mathcal{H}^{(sec,N)}(P,Q,p,q)=\sum_{j=1}^Ng^{(j)}(P,Q,p,q)~~.
\end{equation}
In the model (\ref{hamsum}) we have
\begin{equation}\label{hamsum1}
g^{(1)}(P,Q)=\omega_1 P+\omega_2 Q~~.
\end{equation}
For reasons that will become clear later, for $j\geq2$ we split
each of the functions $g^{(j)}(P,Q,p,q)$ as a sum depending only
on the actions and a sum depending also on the angles:
\begin{equation}\label{hamsum2}
g^{(j)}(P,Q,p,q)=\sum_{\substack{\mathbf{l}\in\mathbb{Z}^2\\
\vert\mathbf{l}\vert=j}}a^{(j)}_{\mathbf{l}}P^{l_1}Q^{l_2}
+\sum_{\substack{\mathbf{l}, \mathbf{k}\in\mathbb{Z}^2\\
\vert\mathbf{l}\vert=j-2}}
b^{(j)}_{\mathbf{l},\mathbf{k}}P^{l_1}Q^{l_2}e^{i(k_1p+k_2q)}~.
\end{equation}

This last splitting completes the second step in the Hamiltonian preparation.
The explicit expressions of the quantities $\omega_1$, $\omega_2$, $a_{\mathbf{l}}$,
$b_{\mathbf{l},\mathbf{k}}$ for $j=2$ are given
in Appendix \ref{appA}, in terms of the orbital elements of the satellite, Moon
and Sun.

\subsection{Preliminary normalization}
\label{ssec norm}
It was already mentioned in Section~\ref{sec intro} that the
presence of the averaged lunisolar terms in \equ{hamsum} implies
the existence of a secular equilibrium solution of Hamilton's
equation's of motion under the Hamiltonian $\mathcal{H}^{(sec)}$,
corresponding to the values $e=0,i=i^{(p)}$ (see Eq.(\ref{ip})),
where $i^{(p)}$ is called the inclination of the \textit{Laplace
plane}. It is easy to see that the non-zero value of the
inclination of the Laplace plane is reflected into the Hamiltonian
$\mathcal{H}^{(sec,N)}$ by the presence of purely trigonometric
terms, i.e., terms with $\vert\mathbf{l}\vert=0$. Such terms yield
coefficients which are dominant with respect to the remaining
terms in the Hamiltonian expansion. Furthermore, in the splitting
of the Hamiltonian as
$\mathcal{H}=h_{0}(\mathbf{I})+h_1(\mathbf{I},\mathbf{u})$, where
$(\mathbf{I},\mathbf{u})$ are action-angle variables, as required
for the implementation of the Nekhoroshev theorem (see next
section), the above terms generate terms with a dominant
coefficient largely affecting the size of the perturbation
$h_1(\mathbf{I},\mathbf{u})$. In the present subsection, we
implement a procedure for controlling the size of the terms
\equ{hamsum} of the expansion, so that we obtain a Hamiltonian
satisfying the norm bounds required for the implementation of the
Nekhoroshev theorem.

More specifically, the aim of the normalization algorithm
described below is to remove, up a certain order $N_{norm}$ with
respect to the expansion (\ref{hamsum}), the angle-dependent terms
which are constant or linear in the actions: this leads to a
Hamiltonian $\mathcal{H}^{(N_{norm})}$, in which the norm of the
angle-dependent part decreases at least quadratically with the
size of the domain $A_{r_0}$ in which local action variables are
defined.

The normalization procedure relies on the use of \textit{Lie series}. In every
normalization step, the transformed Hamiltonian is given by
\begin{equation}
\mathcal{H}^{(new)}=\exp^{(N)}(\mathcal{L}_{\mathcal{\chi}})\mathcal{H}^{(old)},
\end{equation}
where $\mathcal{L}_{\chi}f=\{f,\mathcal{\chi}\}$ ($\{\cdot,\cdot\}$ denotes the
\textit{Poisson bracket}) and $\exp^{(N)}(\mathcal{L}_{\chi})$ is defined by
\begin{equation}
\exp^{(N)}(\mathcal{L}_{\mathcal{\chi}})f=\sum_{s=0}^N\frac{1}{s!}
\mathcal{L}_{\mathcal{\chi}}^sf\ .
\end{equation}

To illustrate the procedure, rename the initial Hamiltonian (\ref{hamsum}) as
$\mathcal{H}^{(0)}$ (where superscripts denote how many normalization steps were
performed). Then:
\begin{equation}\label{hamsum2}
\mathcal{H}^{(0)}(P,Q,p,q)=\sum_{j=1}^Ng^{(j,0)}(P,Q,p,q)\ ,
\end{equation}
where
\begin{eqnarray}\label{abdef}
g^{(1,0)}(P,Q)&=&\omega_1P+\omega_2Q\nonumber\\
g^{(j,0)}(P,Q,p,q)&=&\sum_{\substack{\mathbf{l}\in\mathbb{Z}^2\\\vert\mathbf{l}\vert=j}}
a^{(j,0)}_{\mathbf{l}}P^{l_1}Q^{l_2}
+\sum_{\substack{\mathbf{l},\mathbf{k}\in\mathbb{Z}^2\\\vert\mathbf{l}\vert=j-2}}
b^{(j,0)}_{\mathbf{l},\mathbf{k}}P^{l_1}Q^{l_2}e^{i(k_1p+k_2q)}\, \quad j\geq 2\ .
\end{eqnarray}
The second term of the sum (\ref{hamsum2}) takes the form
\begin{equation}
g^{(2,0)}(P,Q,p,q)=
\sum_{\substack{\mathbf{l}\in\mathbb{Z}^2\\\vert\mathbf{l}\vert=2}}
a^{(2,0)}_{\mathbf{l}}P^{l_1}Q^{l_2}
+
\sum_{ \mathbf{k}\in\mathbb{Z}^2}b^{(2,0)}_{\boldsymbol{0},\mathbf{k}}
e^{i(k_1p+k_2q)}~~.
\end{equation}
The generating function $\chi^{(1)}$ eliminating the above terms has the form
\begin{equation}\label{defchi}
\chi^{(1)}(P,Q,p,q)=
\sum_{\mathbf{l}, \mathbf{k}\in\mathbb{Z}^2}x^{(1)}_{\mathbf{l},\mathbf{k}}
P^{l_1}Q^{l_2}e^{i(k_1p+k_2q)}\ ,
\end{equation}
where the coefficients $x^{(1)}_{\mathbf{l},\mathbf{k}}$ are obtained as
the solution of the homological equation
\begin{equation}
\{\omega_1P+\omega_2Q,\chi^{(1)}\}=
-\sum_{ \mathbf{k}\in\mathbb{Z}^2}b^{(2,0)}_{\boldsymbol{0},\mathbf{k}}
e^{i(k_1p+k_2q)}\ ,
\end{equation}
namely
\begin{equation}
\chi^{(1)}(p,q)=-\sum_{ \mathbf{k}\in\mathbb{Z}^2}
\frac{b^{(2,0)}_{\boldsymbol{0},\mathbf{k}}}{i(\omega_1k_1+\omega_2k_2)}
e^{i(k_1p+k_2q)}\  .
\end{equation}
The normalized Hamiltonian after the first step can be written as
\begin{equation}\label{H2}
\mathcal{H}^{(1)}(P,Q,p,q)=
\omega_1P+\omega_2Q+Z^{(2,1)}(P,Q,p,q)+\sum_{j=3}^Ng^{(j,1)}(P,Q,p,q)\ ,
\end{equation}
where
\begin{equation}
Z^{(2,1)}=g^{(2,0)}+\mathcal{L}_{\chi^{(1)}}(\omega_1P+\omega_2Q)
=\sum_{\substack{\mathbf{l}\in\mathbb{Z}^2\\\vert\mathbf{l}\vert=2}}
a^{(2,0)}_{\mathbf{l}}P^{l_1}Q^{l_2}
\end{equation}
and
\begin{equation}
g^{(j,1)}=\sum_{s=0}^{j-1}\frac{1}{s!}\mathcal{L}_{\chi^{(1)}}^sg^{(j-s,1)}.
\end{equation}
In general, since the generating function $\chi^{(1)}$ is constant in the actions,
one can see that, if $f(P,Q,p,q)$ has polynomial order $\ell$ in the actions, then
the order in the actions of the transformed function $\mathcal{L}_{\chi^{(1)}}f$ is
$\ell-1$. This means that all terms in $\mathcal{H}^{(1)}$ can be labeled through
their polynomial orders in the actions: choosing the expansion order $N$
to be odd and distinguishing the indices
$j$ with respect to their parity, we have, for $n=1,\dots,(N-1)/2$:
\begin{equation}\label{parity}
\begin{split}
g^{(2n,1)}(P,Q,p,q)=
\sum_{\substack{\mathbf{l}\in\mathbb{Z}^2\\\vert\mathbf{l}\vert=2n}}
a^{(2n,1)}_{\mathbf{l}}P^{l_1}Q^{l_2}
+
\sum_{s=0}^{n-1}
\sum_{\substack{\mathbf{l},\mathbf{k}\in\mathbb{Z}^2\\\vert\mathbf{l}\vert=2s}}
b^{(2n,1)}_{\mathbf{l},\mathbf{k}}P^{l_1}Q^{l_2}e^{i(k_1p+k_2q)}\quad (n\geq2), \\
g^{(2n+1,1)}(P,Q,p,q)=
\sum_{\substack{\mathbf{l}\in\mathbb{Z}^2\\\vert\mathbf{l}\vert=2n+1}}
a^{(2n+1,1)}_{\mathbf{l}}P^{l_1}Q^{l_2}
+
\sum_{s=0}^{n-1}
\sum_{\substack{\mathbf{l}, \mathbf{k}\in\mathbb{Z}^2\\\vert\mathbf{l}\vert=2s+1}}
b^{(2n+1,1)}_{\mathbf{l},\mathbf{k}}P^{l_1}Q^{l_2}e^{i(k_1p+k_2q)}.
\end{split}
\end{equation}
After the classical normalization step, the function $Z^{(2,1)}(P,Q,p,q)$ does not
contain angle-dependent terms which are constant or linear in the actions.

The second step focusses on the manipulation of the term
\begin{equation}
g^{(3,1)}(P,Q,p,q)=
\sum_{\substack{\mathbf{l}\in\mathbb{Z}^2\\\vert\mathbf{l}\vert=3}}
a^{(3,1)}_{\mathbf{l},\mathbf{k}}P^{l_1}Q^{l_2}
+
\sum_{\substack{\mathbf{l}, \mathbf{k}\in\mathbb{Z}^2\\\vert\mathbf{l}\vert=1}}
b^{(3,1)}_{\mathbf{l},\mathbf{k}}P^{l_1}Q^{l_2}e^{i(k_1p+k_2q)}\ .
\end{equation}
Precisely, the second normalization step aims to remove the second sum in $g^{(3,1)}$
which is angle-dependent and linear in the actions. The generating function $\chi^{(2)}$,
given by (\ref{defchi}) with a suitable change in the upper indexes, must satisfy the
normal form equations
\begin{equation}
\{\omega_1P+\omega_2Q,\chi^{(2)}\}=
-\sum_{\substack{\mathbf{l}, \mathbf{k}\in\mathbb{Z}^2\\\vert\mathbf{l}\vert=1}}
b^{(3,1)}_{\mathbf{l},\mathbf{k}}P^{l_1}Q^{l_2}e^{i(k_1p+k_2q)}\ ,
\end{equation}
which gives
$$
\chi^{(2)}(P,Q,p,q)= -\sum_{\substack{\mathbf{l},
\mathbf{k}\in\mathbb{Z}^2\\\vert\mathbf{l}\vert=1}}
\frac{b_{\mathbf{l},\mathbf{k}}^{(3,1)}}{i(\omega_1k_1+\omega_2k_2)}P^{l_1}
Q^{l_2}e^{i(k_2p+k_2q)}\ .
$$
As a result, the generating function $\chi^{(2)}$ is linear in the actions, so that
the operator $\mathcal{L}_{\chi^{(2)}}f$ preserves the polynomial degree in the actions
of any generic function $f(P,Q,p,q)$.

The second-order transformed Hamiltonian $\mathcal{H}^{(2)}$ can be written as
\begin{equation}
\mathcal{H}^{(2)}(P,Q,p,q)=\omega_1P+\omega_2Q+\sum_{j=2}^3Z^{(j,2)}(P,Q)
+\sum_{j=4}^Ng^{(j,2)}(P,Q,p,q)\ ,
\end{equation}
where, noticing that $g^{(0,2)}\equiv 0$, one obtains
\begin{equation}
Z^{(2,2)}=\sum_{\substack{\mathbf{l}\in\mathbb{Z}^2\\\vert\mathbf{l}\vert=2}}
a_{\mathbf{l}}^{(2,2)}P^{l_1}Q^{l_2}, \quad
Z^{(3,2)}=\sum_{\substack{\mathbf{l}\in\mathbb{Z}^2\\\vert\mathbf{l}\vert=3}}
a_{\mathbf{l}}^{(3,2)}P^{l_1}Q^{l_2}, \quad
g^{(j,2)}=\sum_{s=0}^{\lfloor\frac{j}{2}\rfloor}\frac{1}{s!}\mathcal{L}^sg^{(j-2s,2)}\ .
\end{equation}
Taking into account the parities of the indexes $j$, one can obtain also for
$g^{(j,2)}$ the analogous of (\ref{parity}).

We can now give the explicit formulas for the normalization steps for $r>2$.
\begin{itemize}
\item
The $r-$th normalization step allows one to transform the Hamiltonian
\begin{equation}
\mathcal{H}^{(r-1)}(P,Q,p,q)=\omega_1P+\omega_2Q
+\sum_{j=2}^{r-1}Z^{(j,r-1)}(P,Q,p,q)+\sum_{j=r}^Ng^{(j,r-1)}(P,Q,p,q)
\end{equation}
into
\begin{equation}
\mathcal{H}^{(r)}(P,Q,p,q)=\omega_1P+\omega_2Q
+\sum_{j=2}^{r}Z^{(j,r)}(P,Q,p,q)+\sum_{j=r+1}^Ng^{(j,r)}(P,Q,p,q),
\end{equation}
with
\begin{eqnarray}\label{generalr}
Z^{(2,r)}&=&
\sum_{\substack{\mathbf{l}\in\mathbb{Z}^2\\\vert\mathbf{l}\vert=2}}
a^{(2,r)}_{\mathbf l}P^{l_1}Q^{l_2},\quad Z^{(3,r)}
=\sum_{\substack{\mathbf{l}\in\mathbb{Z}^2\\\vert\mathbf{l}\vert=3}}
a_{\mathbf{l}}^{(3,r)}P^{l_1}Q^{l_2},\nonumber\\
Z^{(j>3,r)}&=&
\sum_{\substack{\mathbf{l}\in\mathbb{Z}^2\\\vert\mathbf{l}\vert=j}}
a_{\mathbf l}^{(j,r)}P^{l_1}Q^{l_2}
+
\sum_{s=2}^{j-2}
\sum_{\substack{\mathbf{l},\mathbf{k}\in\mathbb{Z}^2\\\vert\mathbf{l}\vert=s}}
b_{\mathbf{l},\mathbf{k}}^{(j,r)}P^{l_1}Q^{l_2}e^{i(k_1p+k_2q)},\nonumber\\
g^{(j,r)}&=&
\sum_{\substack{\mathbf{l}\in\mathbb{Z}^2\\\vert\mathbf{l}\vert=j}}
a_{\mathbf{l}}^{(j,r)}P^{l_1}Q^{l_2}
+
\sum_{s=0}^{j-2}
\sum_{\substack{\mathbf{l},\mathbf{k}\in\mathbb{Z}^2\\\vert\mathbf{l}\vert=s}}
b_{\mathbf{l},\mathbf{k}}^{(j,r)}P^{l_1}Q^{l_2}e^{i(k_1p+k_2q)}~.
\end{eqnarray}
By the above parity rules, which apply also for $r>3$, both $Z^{(j,i)}$ and
$g^{(j,i)}$ contain only the terms with $s$ even if $j$ is even and $s$ odd if $j$ is odd. Notice that, for $j>3$, $Z^{(j,i)}$
can contain also angle-dependent terms, which are at least quadratic in the
actions.
\item
The $r$-th order generating function can be expressed as
\begin{equation}\label{chir}
\chi^{(r)}(P,Q,p,q)=
-\sum_{\substack{\mathbf{l},\mathbf{k}\in\mathbb{Z}^2\\\vert\mathbf{l}\vert=0,1}}
\frac{b_{\mathbf{l},\mathbf{k}}^{(r+1,r-1)}}{i(\omega_1k_1+\omega_2k_2)}
P^{l_1}Q^{l_2}e^{i(k_1p+k_2q)},
\end{equation}
which contains only purely trigonometric terms (independent on the actions) if $r$ is odd and only
terms linear in the actions if $r$ is even.
\item
After $N_{norm}$ normalization steps, the final Hamiltonian is given by
\begin{equation}\label{normham}
\mathcal{H}^{(N_{norm})}(P,Q,p,q)=
\omega_1P+\omega_2Q+\sum_{j=2}^{N_{norm}}Z^{(j,N_{norm})}(P,Q,p,q)
+\sum_{j=N_{norm}+1}^Ng^{(j,N_{norm})}(P,Q,p,q).
\end{equation}
\end{itemize}
From (\ref{generalr}) it is clear that the functions
$g^{(j,\ell)}$ might contain terms which are angle-dependent and
constant or linear in the actions. As we will see later, the
series are convergent in particular domains of the parameters. In
that case, the normalization procedure succeeds to reduce the
magnitude of all the terms in the perturbation to a size
sufficiently small for the application of the Nekhoroshev theorem.

It is also important to observe that particular angle combinations
in the angle-dependent part of the Hamiltonians can produce, if
$r$ is odd, constant terms both in actions and angles, which do
not affect the dynamics; however, when $r$ is even, the same
combinations can produce terms which do not depend on the angles,
but are linear in the actions. These terms represent a
perturbation on the frequencies, which can have important
effects on the applicability of Nekhoroshev theorem.

From the definition of the $r-$th order generating function
(\ref{chir}), one can observe that the convergence of the
normalization algorithm depends heavily on the presence of
\textit{resonances}, which produce small divisors of the type
$\omega_1k_1^{(res)} +\omega_2k_2^{(res)}\approx0$ for suitable
integers $k_1^{(res)}$, $k_2^{(res)}$. Section \ref{sssec conv
prima norm} provides numerical examples of how the presence of
resonances can affect the convergence of the normalization
procedure, along with effects on the variation of the initial
frequencies.

\section{Nekhoroshev stability estimates}
\label{sec prel}

In this Section, we recall the version of the Nekhoroshev theorem
developed in \cite{Poschel} for frequencies satisfying a
non-resonance condition (see  Section~\ref{ssec posch}). Based on
this theorem, we developed an algorithm computing all quantities
needed in order to check whether the necessary conditions for the
holding of the theorem are fulfilled in the case of the
Hamiltonian (\ref{normham}). The algorithm is presented in
Section~\ref{ssec alg}.

\subsection{Theorem on exponential stability}
\label{ssec posch}

Let us consider an $n-$dimensional quasi-integrable Hamiltonian of the form
\begin{equation*}
\mathcal{H}(\mathbf{I},\mathbf{u})=h(\mathbf{I})+f_\epsilon(\mathbf{I},
\mathbf{u})\ ,
\end{equation*}
with $h$ called the integrable part and $f_\epsilon$ the
perturbing function, depending on a small real parameter  $\epsilon>0$. The
Hamiltonian $\mathcal{H}$ is assumed real analytic in the domain
$(\mathbf{I},\mathbf{u})\in A\times\mathbb{T}^n$ with
$A\subseteq\mathbb{R}^n$ open and bounded. Besides, we assume that
$\mathcal{H}$ can be extended analytically to the set $D_{r_0,s_0}$ defined as
\begin{equation}\label{complexneigh}
D_{r_0,s_0}=A_{r_0}\times \mathbb{T}_{s_0}^n\ ,
\end{equation}
where for $r_0,s_0>0$:
\begin{equation}\label{Ar0}
A_{r_0}=\{\mathbf{I}\in\mathbb{C}^n: dist(\mathbf{I}, A)<r_0\}
\end{equation}
and
$$
\mathbb{T}_{s_0}^n=\{\mathbf{u}\in\mathbb{C}^n:Re(u_j)\in\mathbb{T},\
\max_{j=1,\dots,n}\vert Im(u_j)\vert<s_0\}\ .
$$
Finally, we assume that there exists a positive constant $M$ such that
$$
\sup_{\mathbf{I}\in A_{r_0}}\|{\mathcal Q}(\mathbf{I})\|_{o}\leq M\ ,
$$
where ${\mathcal Q}$ denotes the Hessian matrix associated to $h$
and $\|\cdot\|_o$ denotes the operator norm induced by the
Euclidean norm on $\mathbb{R}^n$.

For any analytic function
\begin{equation*}
g(\mathbf{I},
\mathbf{u})=\sum_{\mathbf{k}\in\mathbb{Z}^n}g_{\mathbf{k}}(\mathbf{I})
e^{i\mathbf{k}\cdot\mathbf{u}},
\end{equation*}
in $D_{r_0,s_0}$, we define its Cauchy norm as
\begin{equation}\label{defnorm}
\vert g \vert_{A,r_0,s_0}=\sup_{\mathbf{I}\in
A_{r_0}}\sum_{\mathbf{k}\in\mathbb{Z}^n}\vert
g_{\mathbf{k}}(\mathbf{I})\vert e^{\vert\mathbf{k}\vert s_0}\ ,
\end{equation}
where $|\mathbf{k}|$ is the $\ell^1$-norm of
$\mathbf{k}\in\mathbb{Z}^n$. Finally, let $\epsilon$ be such that
\begin{equation}\label{epsf}
\vert f_\epsilon\vert_{A,r_0,s_0}\leq\epsilon\ .
\end{equation}
The following Theorem provides a bound on the action variables for
exponentially long times; we refer to \cite{Poschel} for the proof
and further extensions. First we need the following definition.

\begin{definition}
A set $D\subseteq A$ is said to be a completely $\alpha,K$-nonresonant domain in $A$,
if for every $\mathbf{k}\in\mathbb{Z}^n\backslash\{\mathbf{0}\}$, $\vert\mathbf{k}\vert\leq K$,
and for every $\mathbf{I}\in D$
\begin{equation}\label{alphadef}
\vert\mathbf{k}\cdot\boldsymbol{\omega}(\mathbf{I})\vert\geq\alpha>0\ ,
\end{equation}
where $\boldsymbol{\omega}(\mathbf{I})=\partial_{\mathbf{I}}h(\mathbf{I})$.
\end{definition}

\begin{theorem}[\cite{Poschel}]\label{posch}
Let $D\subseteq A$ be a completely $\alpha,K$-nonresonant domain.
Let $a,b>0$ such that $\frac{1}{a}+\frac{1}{b}=1$.
Let $\epsilon$ be as in \equ{epsf} for some $r_0,s_0>0$.
If the following inequalities
are satisfied:
\begin{equation}\label{conditions}
\epsilon\leq\frac{1}{2^7b}\frac{\alpha r}{K}=\epsilon^*,
\quad\quad r\leq\min\left(\frac{\alpha}{aMK}, r_0\right)\ ,
\end{equation}
then, denoting by $\vert\vert\cdot\vert\vert$ the Euclidean norm in $A$, one has
\begin{equation}\label{stabtime}
\|\mathbf{I}(t)-\mathbf{I}_0\|\leq r \quad for
\quad \vert t\vert \leq\frac{s_0 r}{5\epsilon}e^{Ks_0/6}
\end{equation}
for every orbit of the perturbed system with initial position $(\mathbf{I}_0,
\mathbf{u}_0)$ in $D\times\mathbb{T}^n$.
\end{theorem}

\subsection{Algorithm for the application of the theorem}
\label{ssec alg}
To apply Theorem~\ref{posch} to the final Hamiltonian $\mathcal{H}^{(N_{norm})}$ defined in
(\ref{normham}), one has to compute all the quantities involved in the Theorem. This procedure
gives rise to an explicit constructive algorithm to give stability estimates for every pair
of reference values $(e_*, i_*)$ in the uniform grid $[0,0.5]\times[0,89.5^{\circ}]$ with
step-size equal to $0.1$ in eccentricity and $0.5^{\circ}$ in inclination. Notice that the
upper value of the grid in inclination is equal to $89.5^{\circ}$ to avoid singularities.

First, we need to determine the greatest integer $\bar{K}$, to which we refer as the
\textit{cut-off value}, such that conditions (\ref{conditions}) hold. From the definition
of $\alpha$ in \equ{alphadef} and $\epsilon^*$ in \equ{conditions}, it is clear that
$\epsilon^*$ decreases as $K$ increases; then, provided that condition (\ref{conditions})
holds for $K=1$, the maximal value $\bar{K}$ exists. On the other hand, if (\ref{conditions})
does not hold for $K=1$, it continues to remain false for all $K>1$.

From a computational point of view, the procedure is composed by the following steps,
(S1), ..., (S8), performed for every pair $(e_*, i_*)$ in the grid defined above; by trial
and error, we fix the values of $r_0$, $s_0$, $a$, $b$. Their choice is arbitrary and can
be tuned so to satisfy the conditions of the Theorem and to optimize the final estimates.

\begin{itemize}
\item[(S1)]
Taylor expansion up to order $N=12$ in the expansion (\ref{hamsum}) around the  actions
$(P_*, Q_*)$, corresponding to the Keplerian elements $(e_*,i_*)$;

\item[(S2)]
normalization up to order $N_{norm}=6$, following the procedure described in
Section~\ref{ssec norm}, which leads to compute the normalized Hamiltonian
$\mathcal{H}^{(N_{norm})}$;

\item[(S3)]
splitting of the Hamiltonian $\mathcal{H}^{(N_{norm})}$ in the unperturbed part $h_0(P,Q)$,
containing the terms of $\mathcal{H}^{(N_{norm})}$ which depend only on the actions,
and the perturbing part $h_1(P, Q, p, q)=\mathcal{H}^{(N_{norm})}(P,Q,p,q)-h_0(P,Q)$;
computation\footnote{With an abuse of notation, we continue to define the new frequencies,
which could be modified by the normalization, with the symbols $\omega_1$ and $\omega_2$.
When, in Section \ref{sssec conv prima norm}, it will be required to distinguish between
the initial and the final frequencies, the latter will be denoted by $\tilde{\omega}_1$
and $\tilde{\omega}_2$.} of $\boldsymbol{\omega}=(\omega_1,\omega_2)$, with $\omega_1$
and $\omega_2$ coefficients respectively of $P$ and $Q$ in $h_0$;

\item[(S4)]
definition of the real and complexified domains in the actions as in (\ref{complexneigh})
and computation of the quantity
\begin{equation}
M=\sup_{(P,Q)\in A_{r_0}}\vert\vert {\mathcal Q}(P,Q)\vert\vert_o\ ;
\end{equation}
in particular, we define $A=[P_*-dP^{(max)},
P_*+dP^{(max)}]\times[Q_*-dQ^{(max)}, Q_*+dQ^{(max)}]$ with $dP^{(max)}=dQ^{(max)}=0.1$;
we select $r_0=s_0=0.1$ and, following \cite{Poschel}, we take $a=9/8$ and $b=9$;

\item[(S5)]
for every $K=1,\dots,50$, computation of the quantities
\begin{equation}\label{alphakdef}
\begin{split}
\alpha_K=\min_{\vert \mathbf{l}\vert\leq
K}{\{\boldsymbol{\omega}\cdot\mathbf{l}\}}, \quad\quad
r_K=\min\left\{\frac{\alpha_K}{a M K},r_0\right\},
\quad\quad\epsilon^*_K=\frac{1}{2^7 b}\frac{\alpha_K r_K}{K}\ ;
\end{split}
\end{equation}

\item[(S6)]
defining $\epsilon=\vert h_1\vert_{A,r_0,s_0}$, check of the condition $\epsilon\leq
\epsilon^*_K$ for every $K=1,\dots,50$;

\item[(S7)]
if $\epsilon\leq \epsilon^*_1$, computation of $\bar{K}$, namely the greatest $K$
such that $\epsilon\leq\epsilon_K^*$,  and of the corresponding stability time
\begin{equation}
t=\frac{s_0 r_{\bar{K}} }{5 \epsilon}e^{\bar{K}s_0/6}\ ;
\end{equation}

\item[(S8)]
if $\epsilon>\epsilon^*_1$, the conditions for the application of Theorem \ref{posch}
are not satisfied. In this case, we impose $\bar{K}=0$.
\end{itemize}

\vskip.1in

We remark that the order of the Taylor expansion $N=12$, the order of normalization
$N_{norm}=6$, the iteration of $K$ up to 50 are set on the basis of having a
reasonable computational execution time on standard laptops.

\section{Results}
\label{ssec numres}
In this Section we present the results of the application of Theorem~\ref{posch} to the
Hamiltonian model described in Section~\ref{sec ham}. This allows us to derive stability
estimates as well as to discuss the convergence of the  normalization procedure.

\subsection{Stability estimates}
\label{sssec main results}
We apply the algorithm of Section~\ref{ssec alg} to probe the Nekhoroshev stability
for satellites with semimajor axes between $11\,000$ km and $19\,000$ km under the model
presented in Section~\ref{sec ham}. The results exposed below highlight the strong
dependence of the stability conditions on the precise values of the elements $(e,i)$.
Of crucial role in this dependence is the location of the `inclination-dependent'
resonances (see Section~\ref{sec intro}). These satisfy a condition of the form $\alpha \dot{p}+\beta\dot{q}=0$
for some coefficients $\alpha$, $\beta\in\integer$.

\begin{table}
\begin{center}
\begin{tabular}{|ccc|ccc|ccc|ccc|}
\hline
 $\alpha$   &$\beta$  &$i(deg)$  &$\alpha$  &$\beta$  &$i(deg)$
&$\alpha$   &$\beta$  &$i(deg)$  &$\alpha$  &$\beta$  &$i(deg)$ \\
\hline
  1         &  0      &  46.37   &   0      &   1     & 90
& 1         &  1      &   0      &   1      &  -1     & 63.4351 \\
  2         &  1      &  33.0156 &  -1      &   2     & 73.1484
&-2         &  1      &  56.0646 &  -2      &   3     & 69.007  \\
 -4         &  3      &  60.0001 &  -4      &   1     & 51.5596
&-1         &  3      &  78.4633 &  -4      &   5     & 66.422  \\
\hline
\end{tabular}
\vskip.1in
\caption{Inclination-dependent resonances which affect the stability in the
lunisolar model. The coefficients $\alpha$ and $\beta$ are such that
$\alpha \dot{p}+\beta\dot{q}= 0$.}
\label{restab}
\end{center}
\end{table}

\vskip.1in

\begin{figure}
\centering
\includegraphics[scale=2.5]{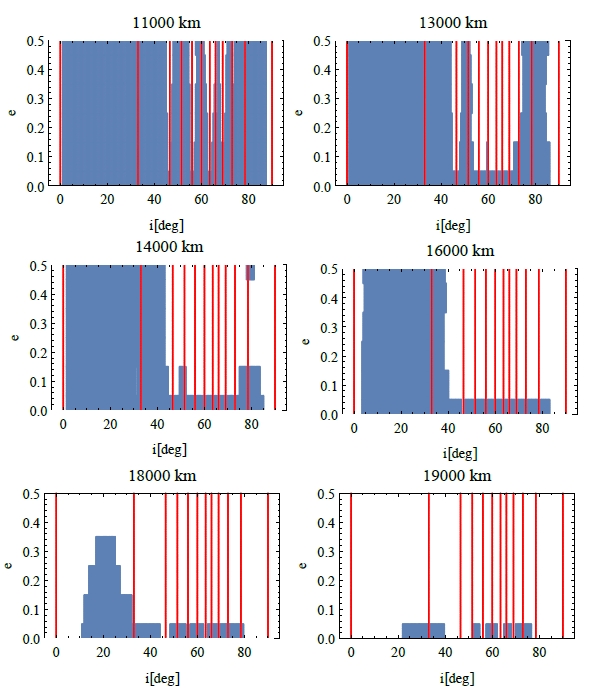}
\caption{Domains of applicability of Theorem~\ref{posch} for different values of the
altitude. The blue regions represent the values of $(i_*,e_*)$ for which the Theorem
can be applied, while the red lines define the values of the inclination which are
associated with the most important resonances in the considered regions
(see Table~\ref{restab}).}
\label{fig:11000-km-mat}
\end{figure}

\begin{figure}
\centering
\includegraphics[scale=3.]{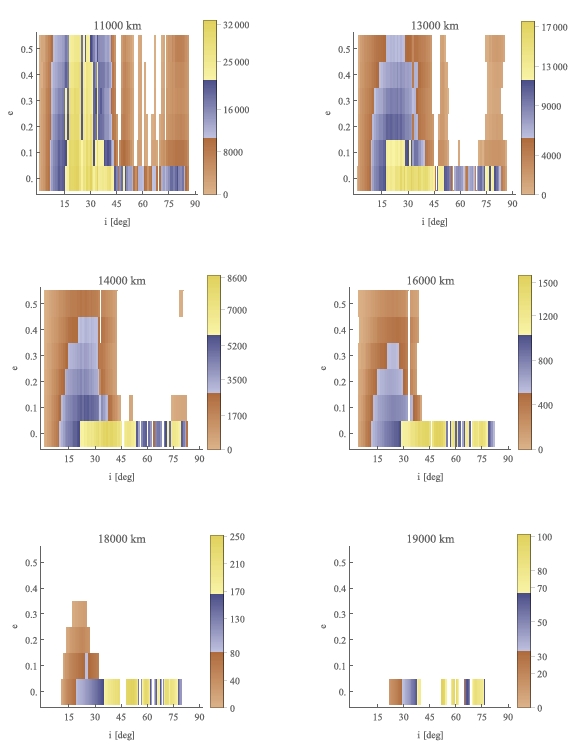}
\caption{Stability time (in years) computed for the values of $(i_*,e_*)$ in the
domain of applicability of Theorem \ref{posch}. }
\label{fig:11000-km-time}
\end{figure}

Table~\ref{restab} shows the values of the inclinations corresponding to each pair of
coefficients $(\alpha,\beta)$. We find that these resonances determine regions where
Theorem~\ref{posch} cannot be applied. This can be exemplified with the help of
Figure \ref{fig:11000-km-mat}, showing (in blue) the region where the algorithm
of Section \ref{ssec alg} returns that the necessary conditions of Theorem
\ref{posch} hold true. The algorithm provides an answer as a function of the chosen
reference values $i_*$ and $e_*$ (for a fixed $a_*$). We take the values of $i_*$ in a
grid by steps of $0.5^\circ$ in the interval $0\leq i_*\leq 89.5^{\circ}$, and of
$e_*$ in a grid by steps of $0.1$ in the interval $0\leq e_*\leq 0.5$. Figure
\ref{fig:11000-km-time} shows the Nekhoroshev stability times computed at every
grid point $(e_*,i_*)$ where the algorithm returns a positive answer for the holding
of the necessary conditions of the theorem.

It is evident from Figure \ref{fig:11000-km-mat} that increasing the distance from the
Earth's center causes a shrinking of the size of the domains of Nekhoroshev stability,
as well as a fast decrease of the corresponding computed stability times. From the
physical point of view, this tendency is evident and can be explained on the basis
of the simple remark that the averaged Hamiltonian $\mathcal{H}_{kep}+\mathcal{H}_{J_2}$ without
third-body perturbations is integrable (the averaged Hamiltonian has no dependence
on the Delaunay angles). Since the overall relative size of third body perturbations
increases with the altitude, these perturbations affect the stability more as $a_*$
increases. At a formal level the effect of the semimajor axis on the estimates
can be identified by an analysis of the convergence of the preliminary normalization
algorithm (see Section~\ref{sssec conv prima norm} below). \\

On the other hand, also evident from Figure \ref{fig:11000-km-mat} is the strong
role of resonances in affecting the stability properties of the system: in fact,
around everyone of the resonances listed in Table \ref{restab} we observe, in the
figure, the formation of a white zone, which indicates values $(e_*,i_*)$ excluded
from the Nekhoroshev stability as detected by our algorithm. As a general comment,
the presence of the resonances acts at two different stages of the computation:
\begin{itemize}
\item[(i)]
it can affect the convergence of the classical normalization, producing an increase
of the size of the perturbing function and a consequent failure of conditions
(\ref{conditions});
\item[(ii)]
near the low-order resonant values of the inclination, the quantity $\alpha_K$
(see (\ref{alphakdef})) can be extremely small, even for low values of $K$. As a
consequence, in the proximity of a resonance, the corresponding value of the
quantity $\epsilon^*_K$ might not be small enough to satisfy (\ref{conditions}).
\end{itemize}


At any rate, we stress that Theorem \ref{posch} used in the present work holds only for
non-resonant domains in the phase space; therefore, by definition it cannot be used to
probe the Nekhoroshev stability very close to resonances. We defer to a future study the
question of the precise investigation of the conditions for Nekhoroshev stability inside
resonances, by implementing a resonant form of the theorem, as first suggested in
\cite{nekhoroshev1977exponential}.

\subsection{Convergence of the preliminary normalization}
\label{sssec conv prima norm}

As pointed out in Section \ref{ssec norm}, the aim of the preliminary normalization
is to allow to control the norm of the perturbing function
$\vert f_\epsilon\vert_{A,r_0,s_0}$ by reducing the size of the complexified action
domain $A_{r_0}$ (see (\ref{Ar0})). In particular, the consequence of the
removal of angle-dependent terms which are constant or linear in the actions is
that, within certain values of the size of the domain $A_{r_0}$, the norm of the
perturbation decreases quadratically with the actions.

\begin{figure}
\centering
\includegraphics[scale=2.]{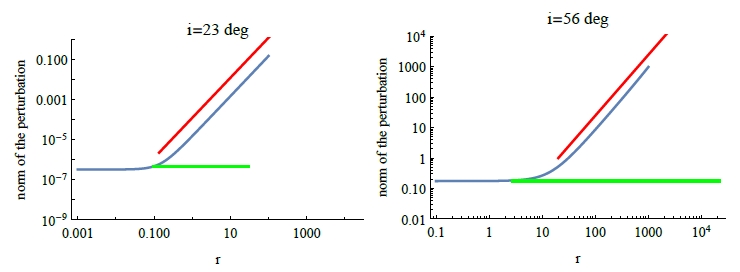}
\caption{Blue: plot in $LogLog$  scale of $\vert
f_\epsilon\vert_{A,r,s_0}$ for $a=13\,000$ km, $e_*=0.2$, and $i_*=23^\circ$ (left)
and $i_*=56^\circ$ (right). For the computations, we selected
$A=[P_*-r,P_*+r]\times[Q_*-r,Q_*+r]$, $r_0=r$, $s_0=0.1$. The slope of the plot
for high $r$ is compared with that of a line with slope $2$ (red); the value at
the plateau (denoted with a  green line) is compared with the value of the norm
of the purely trigonometric part of $f_\epsilon$ with $s_0=0.1$.}
\label{fig:13000-km-quad}
\end{figure}

Figure \ref{fig:13000-km-quad} shows the behaviour of $\vert
f_\epsilon\vert_{A,r,s_0}$ for $a=13\,000$ km, $e_*=0.2$ and two selected values
of $i_*$, as a function of the size of the action in the complexified domain
$A_{r}$ (the domain $A$ is set to be a rectangle of width $2r$ around the central
values $P_*$ and $Q_*$). As expected, the value of $\vert f_\epsilon\vert_{A,r,s_0}$
decreases quadratically with $r$, until it reaches a plateau, whose value is the
norm of the terms of $f_\epsilon$ which do not depend on the actions.

As already mentioned in Section~\ref{sssec main results}, the convergence of the
normalization presented in Section~\ref{ssec norm} for $\mathcal{H}^{(sec)}$ is
crucial to control the size of the perturbing function $h_1$; such value plays a
fundamental role in Theorem~\ref{posch}. A first study of the effect of the
chosen value of the semimajor axis on the convergence can be performed by
considering a simpler model to which a normalization procedure similar to
the one implemented in Section~\ref{ssec norm} can be performed. The
model is defined by the Hamiltonian
\begin{equation}
\tilde{\mathcal{H}}^{(in)}(P,Q,p,q)=\omega_1 P+\omega_2Q +
\frac{c_2}{2}Q^2+ f_1\cos{q}\ ,
\end{equation}
where the frequencies $\omega_1$, $\omega_2$ and the coefficient
$c_2$ depend essentially only on the $J_2$ averaged Hamiltonian,
while the coefficient $f_1$ depends on the lunar and solar
third-body perturbation potentials, and it is proportional to the
sinus of the inclination $i_0$ of the ecliptic. We will examine
the effect of performing the preliminary normalization algorithm
on the Hamiltonian $\tilde{\mathcal{H}}^{(in)}$ so as to remove
purely trigonometric terms. After $N_{norm}$ normalization
steps, the Hamiltonian takes the form:
\begin{equation}
\tilde{\mathcal{H}}^{(fin)}=\omega_1 P+\omega_2Q +
\frac{c_2}{2}Q^2+\sum_{\i=1}^{N_{norm}}
Z_i(P,Q,q)+\sum_{i=N_{norm}+1}^\infty R_i(P,Q,q)\ ,
\end{equation}
where the normalized parts $Z_i(P,Q,q)$ do not contain terms which depend only on
the angle $q$ (as well as linear terms in the actions multiplied by trigonometric
terms). By an explicit computation of the Poisson brackets involved in the normalization,
we readily find that $R_{N_{norm}+1}$ contains trigonometric terms with coefficients
proportional to the quantity
\begin{equation}
2^a f_1 \left(\frac{c_2f_1}{4\omega_2^2}\right)^{N_{norm}},
\end{equation}
where $a=1,2,3$ depends on the value of $N_{norm}$. The convergence of the
remainder through the steps of the normalization algorithm depends, then, on the
value of the ratio $c_2f_1/4\omega_2^2$; in particular, when this quantity is
greater than 1, the normalization does not converge. Neglecting the lunar and solar
contributions in $\omega_1$, $\omega_2$ and $c_2$, the coefficient $c_2f_1/4\omega_2^2$
can be expressed in terms of the orbital elements of debris, Sun and Moon as
\begin{equation}
\frac{c_2f_1}{4\omega_2^2}=\frac{1}{32}\frac{\sin{2i_0}}{R_E^2\mu_EJ_2}
\left(\frac{\mu_M}{(a_M(1-e_M))^3}
+\frac{\mu_{\odot}}{(a_{\odot}(1-e_{\odot}))^3}\right)
a^5(2+3e^2_*)(1-e^2_*)^{3/2}\tan{i_*}~~.
\end{equation}
As a consequence, it is clear that its size strongly depends on $a$ and $i_*$:
it grows sharply when $a$ increases and when $i_*$ approaches $90^\circ$.

On the other hand, the coefficient $f_1$ is proportional to $\sin{2i_0}$, that is, proportional to the (non-zero) inclination $i^{(p)}$ of the Laplace plane (see Eqs. (\ref{ip}) and (\ref{A1,B1})). Hence, the presence in the secular Hamiltonian of purely trigonometric terms is a manifestation of the presence in the model of a Laplace plane. Since $i^{(p)}$ increases with $a$ and $f_1$ increases both with $i^{(p)}$ and $i_*$, this gives a first explanation of the loss of stability of the model as $a$ and
$i_*$ increase.

As already mentioned in Section~\ref{sssec main results}, the other important factor
influencing the size of the remainder across the preliminary normalization process is
the effect of resonances, which, due to Eq.(\ref{chir}), leads to the appearance,
in the series terms, of small divisors. Of particular importance are the small
divisors appearing in the series' purely trigonometric terms, whose size cannot
be controlled by altering the size of the domain in the actions $A_{r_0}$.

\begin{figure}
\centering
\includegraphics[scale=2.]{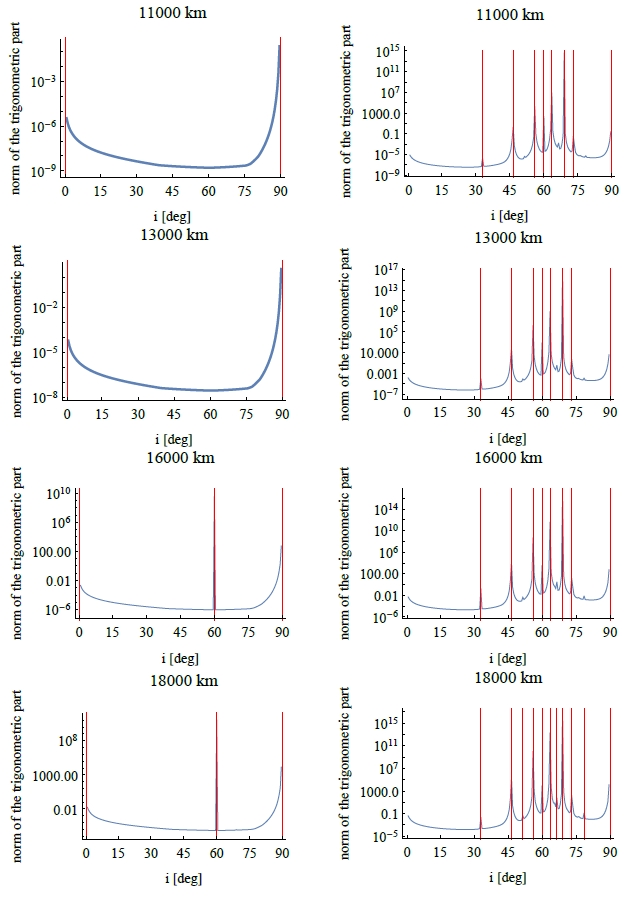}
\caption{Behaviour of the norm of the purely trigonometric part of $h_1$ as a function
of the inclination $i_*$ for different semimajor axes and eccentricities
(left: $e=0$, right: $e=0.5$). The red lines represent the inclinations of the resonances
(see Table \ref{restab}).}
\label{fig:11000-km-e0-trigo}
\end{figure}
Figure \ref{fig:11000-km-e0-trigo} shows the behaviour of the norm of the purely
trigonometric part of the perturbation $h_1$ (with the notation (S3) of
Section~\ref{ssec alg}) as a function of the inclination for four different values of
$a$ and two different values of $e$. As one can see, the size of the trigonometric part
reaches its peaks in correspondence of the resonant values of the inclination, as expected.
We also notice that the number of resonances involved in the growth of the size of the
trigonometric part increases with $a$ and $e$.

As explained in Section~\ref{ssec norm}, the normalization algorithm used in this work
does not perform a re-tuning of the frequencies for every normalization step. This fact
has important effects on the applicability of Theorem~\ref{posch}: when the normalization
converges, the change between the original and the new frequencies is negligible with
respect to their magnitude; on the other hand, when it does not converge, a large
variation in the value of the frequencies occurs, with important consequences
on the computation of $\alpha_K$ and, therefore, of the quantities involved in
Theorem~\ref{posch}.

\begin{figure}
\centering
\includegraphics[scale=3.]{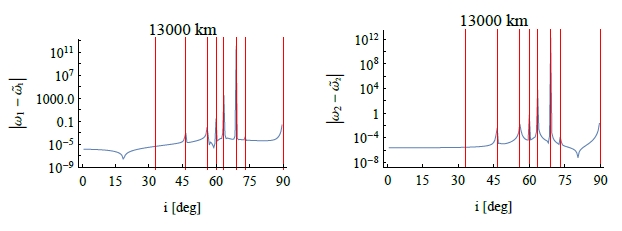}
    \caption{Variation between the initial ($\omega_1$ and $\omega_2$) and
    the final ($\tilde{\omega}_1$ and $\tilde{\omega}_2$)} frequencies as a function of
    the inclination $i_*$ for $a=13\,000$ km and $e_*=0.2$. The red lines
    represent the values of $i_*$ associated to the resonances which affect
    the convergence of the normalization algorithm (see Figure~\ref{fig:11000-km-e0-trigo}).
    \label{fig:13000-km-ome}
 \end{figure}
As an example, Figure~\ref{fig:13000-km-ome} shows the variations of the frequencies as
a function of the inclination for $a=13\,000$ km and $e_*=0.2$.
Comparing Figures~\ref{fig:11000-km-e0-trigo} and \ref{fig:13000-km-ome}, it is clear that
the resonances which affect the growth in size of the purely trigonometric part of $h_1$
and the variation of the frequencies are the same.

\subsection{Behaviour of the cut-off value $\bar{K}$}
\label{sssec cutoff}

\begin{table}
\begin{center}
\begin{tabular}{|ccc|ccc|ccc|ccc|}
\hline
&&&&&&&&&&&\\
 $i_* (deg)$ & $N(i_*)$ & $\bar{K}$ & $i_*(deg)$ & $N(i_*)$ & $\bar{K}$
&$i_* (deg)$ & $N(i_*)$ & $\bar{K}$ &$i_* (deg)$ & $N(i_*)$ & $\bar{K}$   \\
\hline
    46.5     &    1     &     0     &    89.5    &    1     &    0
&    1       &    2     &     1     &    63.5    &    2     &    0        \\
    33       &    3     &     2     &    73      &    3     &    0
&   56       &    3     &     0     &    38      &    3     &    3        \\
    53       &    4     &     1     &    78.5    &    4     &    3
&   40.5     &    4     &     5     &    27      &    5     &    4        \\
    51.5     &    5     &     4     &    58.5    &    5     &    0
&   69       &    5     &     0     &    81.5    &    5     &    4        \\
    41.5     &    6     &     5     &    50.5    &    6     &    5
&   83.5     &    6     &     4     &            &          &             \\
\hline
\end{tabular}
\vskip.1in
\caption{Comparison between the order $N(i_*)$ of the nearest resonance and
the computed cut-off value $\bar{K}$, computed for $a=13\,000$ km and $e=0.1$. }
\label{ktab}
\end{center}
\end{table}

Provided that the classical normalization converges, from the definition of the
cut-off value $\bar{K}$ given in Section~\ref{ssec alg}, one expects that exactly
at a resonance, once denoted with $N(i_*)=\vert\alpha\vert+\vert\beta\vert$ its
order, one has $\bar{K}=N(i_*)-1$. Since in this work the inclinations are
selected in a mesh of $[0,89.5^\circ]$ with step $0.5^\circ$, the computations
of the quantities involved in Theorem~\ref{posch}, including $\bar{K}$, are not
performed exactly at resonance (with the exception of $i_*=60^\circ$, whose distance
from the exact resonance is of the order of $10^{-3}$): Table~\ref{ktab} shows
the value of $\bar{K}$ computed for the points of the mesh which are near to the
resonances up to order $6$, with $a=13\,000$ km and $e=0.1$, along with the resonance
order $N(i_*)$ of the nearest one. With the exception of the
inclinations associated to resonances which affect the convergence of the
classical normalization, the majority of the
listed inclinations follows the expected rule $\bar{K}=N(i_*)-1$, while some slight
deviation is probably due to the numerical computation.

\begin{figure}
\centering
\includegraphics[scale=3.]{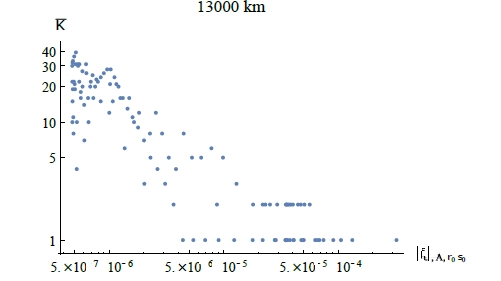}
\caption{Plot in $LogLog$ scale of  the points $\{\vert f_\epsilon\vert_{A,r_0,s_0},
\bar{K}\}$ for $a=13\,000$ km, $e_*=0.2$ and $i_*\in[0,90^\circ]$ on a mesh of step
$0.5^\circ$.}
\label{fig:13000-km-k(eps)}
\end{figure}
To conclude, Figure~\ref{fig:13000-km-k(eps)} shows the relation between the computed values of
$\bar{K}$ and $\vert f_\epsilon\vert_{a,s_0,r_0}$ for $a=13\,000$ km, $e_*=0.2$ and
$i_*\in[0,90^\circ]$. As expected, the cut-off decreases exponentially with the norm
of the perturbing function.

\vskip.1in

\bf Acknowledgements. \rm A.C. partially acknowledges the MIUR Excellence Department Project awarded
to the Department of Mathematics, University of Rome Tor Vergata, CUP
E83C18000100006. A.C. and C.E. were partially supported by the Marie Curie
ITN Stardust-R, GA 813644 of the H2020 research and innovation program.
C.E. acknowledges the MIUR-PRIN 20178CJA2B ``New Frontiers of Celestial
Mechanics: theory and Applications".

\appendix

\section{Analytical expressions of $\mathcal{H}_b^{(av)}$ and $\mathcal{H}^{(sec)}$
in Section~\ref{sec ham}}
\label{appA}

\subsection{Expansion of $\mathcal{H}_b^{(av)}$}
We provide an expression of $\mathcal{H}_b^{(av)}$ for a third body (index $b$, referring to
the Moon or Sun) as a function of its orbital parameters $(a_b,e_b, i_b, \omega_b,\Omega_b)$ and
the debris' parameters $(a,e,i,\omega,\Omega)$. Up to second order in the eccentricity we have:
\begin{equation}\label{Hb}
\begin{split}
&\mathcal{H}_b^{(av)}=\frac{a^2}{16a_b^3(1-e_b^2)^{3/2}}
\Bigg(-\frac{2+3e^2}{8}(1+3\cos{(2i)})(1+3\cos{(2i_0)})+\\
&-\frac{15}{4}e^2(1 + 3 \cos{2i_0}) \sin{i}^2\cos{2 \omega}
-\frac{3}{2} (2 + 3 e^2) \sin{i}^2 \sin{i_0}^2\cos{2( \Omega - \Omega_{b_0})}+\\
&-15 e^2 \cos{(i/2)}^4 \sin{i_0}^2\cos{2( \omega +  \Omega -  \Omega_{b_0})}
-\frac{3}{2} (2 + 3 e^2)  \sin{(2 i)} \sin{(2i_0)}\cos{(\Omega - \Omega_{b_0})} +\\
&+30 e^2 \cos{(i/2)}^3 \sin{(i/2)} \sin{(2 i_0)}\cos{(2 \omega + \Omega - \Omega_{b_0})}\\
&+\frac{15}{2} e^2 (-1 + \cos{i}) \sin{i} \sin{i_0}\cos{(2 \omega - \Omega + \Omega_{b_0})}+\\
&-15 e^2 \sin{(i/2)}^4 \sin{(i_0)}^2\cos{2( \omega - \Omega +  \Omega_{b_0})}
\Bigg)\ .
\end{split}
\end{equation}

\subsection{List of the nonzero terms in $\mathcal{H}^{(sec)}$ for $j=1,2$}
Assuming, as in Section~\ref{sec ham}, that both the lunar and solar orbits lie on
a fixed ecliptic plane inclined with respect to the Earth's equatorial plane by an angle $i_0$,
the frequencies $\omega_1$ and $\omega_2$ appearing in \equ{hamsum1} are given by:
$$
\omega_1=\omega_1^{(J_2)}+\omega_1^{(M)}+\omega_1^{(\odot)}\ ,\qquad\qquad
\omega_2=\omega_2^{(J_2)}+\omega_2^{(M)}+\omega_2^{(\odot)}\ , 
$$
where
\begin{eqnarray}
\omega_1^{(J_2)}&=&
-\frac{3}{4}R_E^2J_2\mu_E^4
\frac{(-1+5\cos{i_*}^2-2\cos{i_*})}{(\mu_E a)^{7/2}(1-e_*^2)^2}\nonumber\\
\omega_2^{(J_2)}&=&
\frac{3}{2}\frac{R_E^2J_2\mu_E^4}{(\mu_E a)^{7/2}(1-e^2)^2}\cos{i_*}\nonumber\\
\omega_1^{(M/\odot)}&=&
-\frac{3}{64}a^{3/2}\mu_{M/\odot}
\frac{[3+2e_*^2-2(2+3e_*^2)\cos{i_*}
+5\cos{2i_*}](1+3\cos{2i_0})}{\sqrt{1-e_*^2}\sqrt{\mu_E}(a_{M/\odot}(1-e_{M/\odot}))^3}\nonumber\\
\omega_2^{(M/\odot)}& =&\frac{3}{32}a^{3/2}\mu_{M/\odot}\frac{\left(2+3e_*^2\right)\cos{i_*}\left(1+3\cos{2i_0}\right)}{\sqrt{1-e_*^2}\sqrt{\mu_E}\left(a_{M/\odot}(1-e_{M/\odot})\right)^3}.
\end{eqnarray}
The coefficients $a_{\mathbf{l}}$ and $b_{\mathbf{l},\mathbf{k}}$ in \equ{abdef} are
given by: 
\begin{equation*}
\begin{split}
a_{(2,0)}=&\frac{3}{4}\frac{J_2R_E^2}{a^4(1-e_*^2)^{5/2}}(1+10\cos{i_*}-15\cos^2{i_*})+\\
&-\frac{3}{128}\frac{a}{\mu_E(1-e_*^2)}\left(\frac{\mu_M}{R_M^3}+\frac{\mu_{\odot}}{R_{\odot}^3}\right)(1+3\cos{2i_0})(21+4e_*^2-40\cos{i_*}+15\cos{2i_*})\\
a_{(1,1)}=&\frac{3}{2}\frac{J_2R_E^2}{a^4(1-e_*^2)^{5/2}}(5\cos{i_*}-1)+\\
&-\frac{3}{32}\frac{a}{\mu_E(1-e_*^2)}\left(\frac{\mu_M}{R_M^3}+\frac{\mu_{\odot}}{R_{\odot}^3}\right)(1+3\cos{2i_0})(2+3e_*^2-10\cos{i_*})\\
a_{(0,2)}=&-\frac{3}{4}\frac{J_2R_E^2}{a^4(1-e_*^2)^{5/2}}+\\
&-\frac{3}{64}\frac{a}{\mu_E(1-e_*^2)}\left(\frac{\mu_M}{R_M^3}+\frac{\mu_{\odot}}{R_{\odot}^3}\right)(1+3\cos{2i_0})\left(2+3e_*^2\right)
\end{split}
\end{equation*}

\begin{equation*}
\begin{split}
&b_{(0,0),(\pm2,0)}=-\frac{15}{32}(a^2e_*^2\sin^{2}{i_0}\cos^4{(i_*/2)})\left(\frac{\mu_M}{r_M^3}+\frac{\mu_{\odot}}{r_{\odot}^3}\right)\\
&b_{(0,0),(\pm2,\pm1)}=\frac{15}{16}\left[a^2e_*^2\sin{(2i_0)}\cos^3{(i_*/2)}\sin{(i_*/2)}\right]\left(\frac{\mu_M}{r_M^3}+\frac{\mu_{\odot}}{r_{\odot}^3}\right)\\
&b_{(0,0),(\pm2,\mp2)}=-\frac{15}{128}\left[a^2e_*^2(i+3\cos{(2i_0)})\sin^2{i_*}\right]\left(\frac{\mu_M}{r_M^3}+\frac{\mu_{\odot}}{r_{\odot}^3}\right)\\
&b_{(0,0),(\pm2,\mp3)}=-\frac{15}{16}\left[a^2e_*^2\sin{(2i_0)}\sin^3{(i_*/2)}\cos{(i_*/2)}\right]\left(\frac{\mu_M}{r_M^3}+\frac{\mu_{\odot}}{r_{\odot}^3}\right)\\
&b_{(0,0),(\pm2,\mp4)}=-\frac{15}{32}\left[a^2e_*^2\sin^2{i_0}\sin^4{(i_*/2)}\right]\left(\frac{\mu_M}{r_M^3}+\frac{\mu_{\odot}}{r_{\odot}^3}\right)\\
&b_{(0,0),(0,\pm1)}=-\frac{3}{64}\left[a^2(2+3e_*^2)\sin{(2i_0)}\cos{(2i_*)}\right]\left(\frac{\mu_M}{r_M^3}+\frac{\mu_{\odot}}{r_{\odot}^3}\right)\\
&b_{(0,0),(0,\pm2)}=-\frac{3}{64}\left[a^2(2+3e_*^2)\sin^2{i_0}\sin^2{i_*}\right]\left(\frac{\mu_M}{r_M^3}+
\frac{\mu_{\odot}}{r_{\odot}^3}\right)\ .
\end{split}
\end{equation*}

\bibliographystyle{alpha}
\bibliography{bibliografia}

\newcommand{\etalchar}[1]{$^{#1}$}
\begin{thebibliography}{GDGR16}

\bibitem[ADR{\etalchar{+}}16]{ADRRVDQM}
E.~M. Alessi, F.~Deleflie, A.J. Rosengren, A.~Rossi, G.B. Valsecchi, J.~Daquin,
  and K.~Merz.
\newblock A numerical investigation on the eccentricity growth of {GNSS}
  disposal orbits.
\newblock {\em Celest. Mech. Dyn. Astr.}, 125(1):71--90, 2016.

\bibitem[AHA21]{Aristoff2021}
J.M. Aristoff, J.T. Horwood, and K.T. Alfriend.
\newblock On a set of {$J_2$} equinoctial orbital elements and their use for
  uncertainty propagation.
\newblock {\em Celest. Mech. Dyn. Astr.}, 133(9), 2021.

\bibitem[Arn64]{Arnold64}
V.I. Arnold.
\newblock Instability of dynamical systems with several degrees of freedom.
\newblock {\em Sov. Math. Doklady}, 5:581--585, 1964.

\bibitem[BG86]{benettin1986stability}
G.~Benettin and G.~Gallavotti.
\newblock Stability of motions near resonances in quasi-integrable
  {H}amiltonian systems.
\newblock {\em Journal of Statistical Physics}, 44(3-4):293--338, 1986.

\bibitem[Bre01a]{sB01}
S.~Breiter.
\newblock Lunisolar resonances revisited.
\newblock {\em Celest. Mech. Dyn. Astr.}, 81:81--91, 2001.

\bibitem[Bre01b]{breiter2001coupling}
S.~Breiter.
\newblock On the coupling of lunisolar resonances for {E}arth satellite orbits.
\newblock {\em Celestial Mechanics and Dynamical Astronomy}, 80(1):1--20, 2001.

\bibitem[CEG{\etalchar{+}}17]{celletti2017dynamical}
A.~Celletti, C.~Efthymiopoulos, F.~Gachet, C.~Gale{\c{s}}, and G.~Pucacco.
\newblock Dynamical models and the onset of chaos in space debris.
\newblock {\em International Journal of Non-Linear Mechanics}, 90:147--163,
  2017.

\bibitem[Cel10]{celletti2010stability}
A.~Celletti.
\newblock {\em Stability and Chaos in Celestial Mechanics}.
\newblock Springer-Verlag, Berlin; published in association with Praxis
  Publishing, Chichester, 2010.

\bibitem[CF96]{CF}
A.~Celletti and L.~Ferrara.
\newblock An application of {N}ekhoroshev theorem to the restricted three-body
  problem.
\newblock {\em Celest. Mech. Dyn. Astr.}, 64:261--272, 1996.

\bibitem[CG91]{CG}
A.~Celletti and A.~Giorgilli.
\newblock On the stability of the {L}agrangian points in the spatial restricted
  problem of three bodies.
\newblock {\em Cel. Mech. Dyn. Astr.}, 50:31--58, 1991.

\bibitem[CG04]{chaogick}
C.C. Chao and R.A. Gick.
\newblock Long-term evolution of navigation satellite orbits:
  Gps/glonass/galileo.
\newblock {\em Advances in Space Research}, 34:1221--1226, 2004.

\bibitem[CG14]{CGmajor}
A.~Celletti and C.~Gales.
\newblock On the dynamics of space debris: 1:1 and 2:1 resonances.
\newblock {\em J. Nonlinear Science}, 24(6):1231--1262, 2014.

\bibitem[CG18]{celletti2018dynamics}
A.~Celletti and C.~Gale{\c{s}}.
\newblock Dynamics of resonances and equilibria of low {E}arth objects.
\newblock {\em SIAM Journal on Applied Dynamical Systems}, 17(1):203--235,
  2018.

\bibitem[CGL20]{CGL2020}
A.~Celletti, C.~Gales, and C.~Lhotka.
\newblock Resonances in the {E}arth's space environment.
\newblock {\em Comm. Nonlinear Sciences and Numerical Simulations}, 84:105185,
  2020.

\bibitem[CGP16]{CGPSIADS}
A.~Celletti, C.~Gales, and G.~Pucacco.
\newblock Bifurcation of lunisolar secular resonances for space debris orbits.
\newblock {\em SIAM J. Appl. Dyn. Syst.}, 15:1352--1383, 2016.

\bibitem[CGPR17]{CGPR2017}
A.~Celletti, C.~Gales, G.~Pucacco, and A.~Rosengren.
\newblock Analytical development of the lunisolar disturbing function and the
  critical inclination secular resonance.
\newblock {\em Celest. Mech. Dyn. Astron.}, 127(3):259--283, 2017.

\bibitem[Coo62]{cook1962luni}
G.~E. Cook.
\newblock Luni-solar perturbations of the orbit of an {E}arth satellite.
\newblock {\em Geophysical Journal International}, 6(3):271--291, 1962.

\bibitem[CPL15]{CPL2015}
D.~Casanova, A.~Petit, and A.~Lema\^{\i}tre.
\newblock Long-term evolution of space debris under the {$J_2$} effect, the
  solar radiation pressure and the solar and lunar perturbations.
\newblock {\em Celest. Mech. Dyn. Astr.}, 123:223--238, 2015.

\bibitem[DBCE21]{DCE}
I.~De~Blasi, A.~Celletti, and C.~Efthymiopoulos.
\newblock Semi-analyitical estimates for the orbital stability of {E}arth's
  satellite.
\newblock {\em J. Nonlinear Science}, 31(93), 2021.

\bibitem[DRA{\etalchar{+}}16]{daquin2016dynamical}
J.~Daquin, A.J. Rosengren, E.M. Alessi, F.~Deleflie, G.B. Valsecchi, and
  A.~Rossi.
\newblock The dynamical structure of the {MEO} region: long-term stability,
  chaos, and transport.
\newblock {\em Celestial Mechanics and Dynamical Astronomy}, 124(4):335--366,
  2016.

\bibitem[EH97]{EH}
T.A. Ely and K.C. Howell.
\newblock Dynamics of artificial satellite orbits with tesseral resonances
  including the effects of luni-solar perturbations.
\newblock {\em Dynamics and Stability of Systems}, 12(4):243--269, 1997.

\bibitem[GDGR16]{gkolias2016order}
I.~Gkolias, J.~Daquin, F.~Gachet, and A.J. Rosengren.
\newblock From order to chaos in {E}arth satellite orbits.
\newblock {\em The Astronomical Journal}, 152(5):119, 2016.

\bibitem[Gia74]{giacaglia}
G.E.O. Giacaglia.
\newblock Lunar perturbations of artificial satellites of the {E}arth.
\newblock {\em Celest. Mech.}, 9:239--267, 1974.

\bibitem[GS97]{G2}
A.~Giorgilli and C.~Skokos.
\newblock On the stability of the {T}rojan asteroids.
\newblock {\em Astron. Astrophys.}, 317:254--261, 1997.

\bibitem[Hug80]{Hughes}
S.~Hughes.
\newblock {E}arth satellite orbits with resonant lunisolar perturbations. i.
  {R}esonances dependent only on inclination.
\newblock {\em Proc. R. Soc. Lond. A}, 372:243--264, 1980.

\bibitem[Kau62]{Kaula1962}
W.M. Kaula.
\newblock Development of the lunar and solar disturbing functions for a close
  satellite.
\newblock {\em Astron. J.}, 67:300--303, 1962.

\bibitem[Lan89]{Lane1989}
M.~T. Lane.
\newblock On analytic modeling of lunar perturbations of artificial satellites
  of the {E}arth.
\newblock {\em Celest. Mech. Dynam. Astr.}, 46(4):287--305, 1989.

\bibitem[LCG16]{Lho2016}
C.~Lhotka, A.~Celletti, and C.~Gales.
\newblock {P}oynting-{R}obertson drag and solar wind in the space debris
  problem.
\newblock {\em Mon. Not. Roy. Ast. Soc.}, 460:802--815, 2016.

\bibitem[LDV09]{LDV}
A.~Lema\^{\i}tre, N.~Delsate, and S.~Valk.
\newblock A web of secondary resonances for large {$A/m$} geostationary debris.
\newblock {\em Celest. Mech. Dyn. Astr.}, 104:383--402, 2009.

\bibitem[Nek77]{nekhoroshev1977exponential}
N.N. Nekhoroshev.
\newblock An exponential estimate of the time of stability of nearly-integrable
  {H}amiltonian systems.
\newblock {\em Uspekhi Matematicheskikh Nauk}, 32(6):5--66, 1977.

\bibitem[NG21]{Nie2021}
T.~Nie and P.~Gurfil.
\newblock Long-term evolution of orbital inclination due to third-body
  inclination.
\newblock {\em Celest. Mech. Dyn. Astr.}, 133(1), 2021.

\bibitem[P{\"o}s93]{Poschel}
J.~P{\"o}schel.
\newblock Nekhoroshev estimates for quasi-convex hamiltonian systems.
\newblock {\em Mathematische Zeitschrift}, 213(1):187--216, 04 1993.

\bibitem[RARV15]{aR15}
A.J. Rosengren, E.M. Alessi, A.~Rossi, and G.B. Valsecchi.
\newblock Chaos in navigation satellite orbits caused by the perturbed motion
  of the moon.
\newblock {\em Mon. Not. R. Astron. Soc.}, 449:3522--3526, 2015.

\bibitem[RDT{\etalchar{+}}16]{rosengren2016galileo}
A.J. Rosengren, J.~Daquin, K.~Tsiganis, E.M. Alessi, F.~Deleflie, A.~Rossi, and
  G.B. Valsecchi.
\newblock Galileo disposal strategy: stability, chaos and predictability.
\newblock {\em Monthly Notices of the Royal Astronomical Society},
  464(4):4063--4076, 2016.

\bibitem[Ros08]{rossi2008resonant}
A.~Rossi.
\newblock Resonant dynamics of medium {E}arth orbits: space debris issues.
\newblock {\em Celestial Mechanics and Dynamical Astronomy}, 100(4):267--286,
  2008.

\bibitem[RS13]{Rosengren2013}
A.J. Rosengren and D.J. Scheeres.
\newblock Long-term dynamics of high area-to-mass ratio objects in high-{E}arth
  orbit.
\newblock {\em Adv. Space Res.}, 52:1545--1560, 2013.

\bibitem[SRTV19]{skoulidou2019medium}
D.K. Skoulidou, A.J. Rosengren, K.~Tsiganis, and G.~Voyatzis.
\newblock Medium {E}arth orbit dynamical survey and its use in passive debris
  removal.
\newblock {\em Advances in Space Research}, 63(11):3646--3674, 2019.

\bibitem[VL08]{VL}
S.~Valk and A.~Lema\^{\i}tre.
\newblock Analytical and semi-analytical investigations of geosynchronous space
  debris with high area-to-mass ratios.
\newblock {\em Advances in Space Research}, 41:1077--1090, 2008.

\end{thebibliography}

\end{document}